\documentclass{rspublic}
\usepackage{amssymb,amsmath}

\begin{document}

\title[Integrability and Linearization]{On the complete integrability and
linearization of nonlinear
ordinary differential equations - Part V: Linearization of coupled second order equations}

\author[Chandrasekar, Senthilvelan and Lakshmanan]{V. K. Chandrasekar,
M. Senthilvelan and M. Lakshmanan}

\affiliation{Centre for Nonlinear Dynamics, Department of Physics,
Bharathidasan Univeristy, Tiruchirapalli - 620 024, India}

\label{firstpage}

\maketitle

\begin{abstract}{Nonlinear differential equations, Coupled second order,
Integrability, Linearization}
Linearization of coupled second order nonlinear ordinary
differential equations (SNODEs) is one of the open and challenging
problems in the theory of differential equations. In this paper we
describe a simple and straightforward method to derive linearizing
transformations for a class of two coupled SNODEs.  Our procedure
gives several new types of linearizing transformations of both
invertible and non-invertible kinds.  In both the cases we
provide algorithms to derive the general solution of the given
SNODE. We illustrate the theory with potentially important
examples.
\end{abstract}

\section{Introduction}
Continuing our study on the integrability and linearization of
coupled second order nonlinear ordinary differential equations (SNODEs), in
the present paper, we focus our attention on the linearization of two
coupled SNODEs.  The present study arises not only for the completeness
of Part IV (Chandrasekar \textit{et al.} 2009)
but also to show the importance of unfinished tasks that exist in
the theory of linearization of two coupled SNODEs. As
far as the first point is concerned, we
show that one can also solve a class of coupled second order nonlinear ODEs by transforming them
into two second order free particle equations and from the
solutions of the latter one can construct the solution of the former,
even though this is a nontrivial problem in many situations (one can
also transform coupled nonlinear ODEs into uncoupled nonlinear ones,
which has already been pointed out by us in the previous paper, Part
IV). Regarding the second point, we wish to stress the fact that
linearization of coupled nonlinear ODEs is a vast area of research
which is still in its early stage. In this paper, we show that in
spite of the difficulties that exist in this topic, one can make useful
progress on certain issues, namely (i) developing a method to
deduce all linearizing transformations wherein the new dependent
variables are functions of the old dependent and independent
variables only and not derivatives of the dependent variables, and (ii)
developing a method of
constructing solutions of nonlinear ODEs from the linear ones in the
case of non-point transformations.


Even though the modern theory of linearization of nonlinear ODEs had
originated and developed with the works of Lie, Tresse and Cartan
(Mahomed \& Leach 1989; Steeb 1993; Olver 1995; Ibragimov 1999;
 Chandrasekar \textit{et al.} 2005), the entire
subject was lying dormant for more than a century. Only recently, during
the past two decades or so, notable progress has been made to
linearize nonlinear ODEs through non-point (Duarte \textit{et al.}
1994) or generalized transformations (Chandrasekar \textit{et al.}
2006). For example, focussing our attention on single second
order ODEs, generalized Sundmann (Euler \textit{et al.} 2003; Euler
\textit{et al.} 2004) and generalized linearizing transformations
(Chandrasekar \textit{et al.} 2006) have been introduced to
linearize a class of equations that cannot be linearized by
invertible point transformations. As far as two coupled
SNODEs are concerned, to our knowledge, most of the studies
were focussed only on invertible point transformations
irrespective of whether it
is an analytical approach or a geometrical formulation. For a
survey on this topic one may refer the recent papers of
Merker (2006) and Mahomed \& Qadir (2007) and for the earlier works 
in this direction we cite
Crampin \textit{et al.} (1996), Fels (1995), Grossman
(2000), Soh \& Mahomed  (2001) and Qadir (2007). 
More recently Sookmee \& Meleshko (2008) proposed a new algorithm to linearize the coupled second order ODEs by sequentially reducing the order of the equation. 



In this work we aim to give a new dimension to the theoretical
development of  linearization of nonlinear dynamical
systems having two degrees of freedom by proving that one can unearth a wide class of linearizing
transformations besides invertible point transformations.  Of course,
the latter ones form a sub-class of the new ones which we construct
in this paper. In the present study, we not only derive several new
types of linearizing
transformations but also propose  systematic procedures to derive
the general solution in all these cases.  We also wish to emphasize
here that we derive all these transformations from the first two
integrals alone and thereby establish a potentially simple,
straightforward and powerful approach in the theory of differential
equations.

The plan of the paper is as follows. In \S2$\;a$ we briefly describe the
method of deriving linearizing transformations for a system of two coupled
second-order ODEs. We show that one can have two classes of linearizing transformations, depending upon the nature of the independent variables. If the new independent variables are the same
($z_1=z_2$), we put them in Class-A category, and if they are different ($z_1\neq z_2$), then we put them in Class-B category. In \S2$\;b(i)$ we consider Class-A category and identify three types of linearizing transformations. In \S2$\;b(ii)$ we consider Class-B category and identify six types of linearizing transformations. In \S3 we consider one specific example for each of the nine types of linearizing transformations we have identified, and obtain general solution to each one of them to demonstrate our procedure. Finally, we present our conclusions in \S4.

\section{Linearizing transformations}
\subsection{Method of deriving linearizing transformations}
To begin with let us consider a system of two coupled second order nonlinear ODEs,
$\it{R}\{t,\bf{x}\}$, (vide equation (2.1) in Part IV (Chandrasekar \textit{et al.} 2009)),
\begin{eqnarray}
\ddot{x}=\phi_1(t,x,y,\dot{x},\dot{y}),\quad \mbox{and}\quad
\ddot{y}=\phi_2(t,x,y,\dot{x},\dot{y}). \label{eq01}
\end{eqnarray}
Any transformation of the form $\it{T}\{t,\bf{x}\}$ defined by
\begin{align}
w_1 = f_1(t,x,y),\quad z_1 = \int f_3(t,x,y,\dot{x},\dot{y})
dt,
\nonumber\\
w_2 = f_2(t,x,y),\quad z_2 = \int f_4(t,x,y,\dot{x},\dot{y})
dt, \label{lintra01}
\end{align}
which transforms the given set of nonlinear ODEs (\ref{eq01}) to the free particle equations,
\begin{eqnarray}
\frac {d^2w_1}{dz_1^2}=0,\qquad \frac {d^2w_2}{dz_2^2}=0,
\label{met13h}
\end{eqnarray}
is called a linearizing transformation in the present work.

Let
\begin{eqnarray}
I_1=\mathcal{F}(t,x,y,\dot{x},\dot{y}),\quad \mbox{and}\quad
I_2=\mathcal{G}(t,x,y,\dot{x},\dot{y}). \label{met13aa}
\end{eqnarray}
be the two first integrals of motion of the coupled system (\ref{eq01}) and that they can be explicitly found, if they exist, for example, by using generalized modified Prelle-Singer (PS) method formulated in Paper IV (Chandrasekar \textit{et al.} 2009). Then the following theorem ensures that the transformation can be deduced from the $I_i$'s, $i=1,2$.
\vskip .2cm
\noindent {\it {{\bf Theorem:}
Suppose a given nonlinear system $\it{R}\{t,\bf{x}\}$ of ODEs (\ref{eq01}) is linearizable to a system of two uncoupled free particle equations through the linearizing transformation $\it{T}\{t,\bf{x}\}$ of the form (\ref{lintra01}), then the latter can be deduced from the first integrals $I_i(t,x,y,\dot{x},\dot{y})$, $i=1,2$.}}
\vskip .2cm
\noindent {\it{Proof:}} Let us re-express each of the functions $\mathcal{F}$ and $\mathcal{G}$ in 
(\ref{met13aa}) as a product of two new functions, that is,
\begin{eqnarray}
&&I_1=\frac{1}{f_3(t,x,y,\dot{x},\dot{y})}\frac{d}{dt}f_1(t,x,y)
,\nonumber\\
&&I_2=\frac{1}{f_4(t,x,y,\dot{x},\dot{y})}\frac{d}{dt}f_2(t,x,y).
\label{met13b}
\end{eqnarray}
Again rewriting $f_3$ and $f_4$ as total time derivatives of another
set of functions, say $z_1$ and $z_2$, respectively, that is
$dz_1/dt=f_3(t,x,y,\dot{x},\dot{y})$ and
$dz_2/dt=f_4(t,x,y,\dot{x},\dot{y})$, equation (\ref{met13b}) can be
further recast as
\begin{eqnarray}
I_1=\frac{1}{\frac{dz_1}{dt}}\frac{df_1}{dt}
=\frac
{df_1}{dz_1},\quad
I_2=\frac{1}{\frac{dz_2}{dt}}\frac{df_2}{dt}=\frac
{df_2}{dz_2}. \label{met13d}
\end{eqnarray}

Now identifying the functions $f_1(t,x,y)=w_1$ and $f_2(t,x,y)=w_2$
as the new dependent variables, equation (\ref{met13d}) can be further 
recast in the form
\begin{eqnarray}
\frac {dw_1}{dz_1}=\hat{I_1}, \quad \frac {dw_2}{dz_2}=\hat{I_2},
\label{met13g}
\end{eqnarray}
where $\hat{I}_1$ and $\hat{I}_2$ are the redefined constants.  Obviously, (\ref{met13h}) follows straightforwardly from (\ref{met13g}). Consequently the new variables, $z_i$ and $w_i$, $i=1,2$, defined by equation (\ref{lintra01})
help us to transform the given set of coupled SNODEs into two linear
second order ODEs which in turn lead to the solution by trivial
integration. The variables $w_i$'s and $z_i$'s, $i=1,2,$ then define
the linearizing transformations for the given equation (\ref{eq01}).
\hspace{9cm}$\Box$

It may be noted that in general the new dependent variables, $w_1$ and 
$w_2$, may also involve $\dot{x}$ and $\dot{y}$, that is  $w_1 =
G_1(t,x,y,\dot{x},\dot{y})$ and $w_2=G_3(t,x,y,\dot{x},\dot{y})$, and
this possibility may lead us to identify more generalized transformations like
point-contact and generalized-contact transformations. 
However, in this paper we will confine ourselves only to the forms
of $w_1$ and $w_2$ given by (\ref{lintra01}). 

\subsection{The nature of transformations}
An important question which we will focus upon in the present paper is that
what are the possible forms of linearizing transformations one can
unearth through the above procedure. We recall here
that  in the case of scalar SNODEs, one has
point, generalized Sundman and generalized linearizing
transformations (Chandrasekar \textit{et al.} 2005;2006).
As far as the coupled SNODEs (\ref{eq01}) are concerned, since there are two independent
variables $z_1$  and $z_2$ as given in equation (\ref{lintra01}),
one can choose them to be either the same, $z_1=z_2$ (Class-A), or  different, $z_1\neq z_2$ (Class-B).
In the case of Class-A transformations, one can construct three
different types of linearizing transformations, while for class-B
one can formulate six different types of linearizing
transformations, as we point out below. However, we also note that
even further types of local transformations involving the variables
$\dot{x}$ and $\dot{y}$ are possible but these are not included in
the present study.

\subsubsection{Class-A Linearizing Transformations ($z_1=z_2=z$)}
\label{sec5:7:2:1} In the case of Class A transformations we have $w_1 = f_{1}(t,x,y),
\;w_2 = f_{2}(t,x,y)$, $z_1 =z_2 =z= \int
f_3(t, x, y,\dot{x}, \dot{y}) dt = \int
f_4(t, x, y,$ $ \dot{x}, \dot{y}) dt$. Now appropriately restricting the
form of $f_3$ $(=f_4)$, one can identify three different types of linearizing
transformations.
\begin{enumerate}
\item{
Suppose $z_1=z_2=z$ is a perfect
differential function and $w_i$'s, $i=1,2,$ and $z$ do not
contain the variables $\dot{x}$ and $\dot{y}$, then we call the
resultant transformation, namely, $w_1=f_1(t,x,y)$, $w_2=f_2(t,x,y)$,
$z=f_3(t,x,y)$, as a point transformation of type-I.}
\item{ On
the other hand, if $z$ is not a perfect differential
function, and  $w_i$'s, $i=1,2,$ and $z$ do not contain the
variables $\dot{x}$ and $\dot{y}$, then we call the resultant
transformation, namely, $w_1=f_1(t,x,y)$, $w_2=f_2(t,x,y)$,
$z=\int f_3(t,x,y)dt$, as a generalized Sundman transformation
of type-I.}
\item{ As a more general case, if we consider the independent
variable $z$ to
contain the derivative terms also, that is $w_1=f_1(t,x,y)$,
$w_2=f_2(t,x,y)$, $z=\int f_3(t,x,y,\dot{x},\dot{y})dt$, then we
call the resultant transformation as a generalized linearizing
transformation of type-I.}
\end{enumerate}
In our analysis, we do not consider the possibility $w_1=f_1(t,x,y)$,
$w_2=f_2(t,x,y)$ and $z=f_3(t,x,y,\dot{x},\dot{y})$ because the
procedure to handle it is different from the presently discussed
linearizing transformations. This possibility will be studied separately.
\subsubsection{Class-B Linearizing Transformations ($z_1\neq z_2$)}
\label{sec5:7:2:2} In the Class-B type of linearizing transformations we have $w_1 =
f_{1}(t,x,y), \;w_2 = f_{2}(t,x,y)$ and $z_1 =\int
f_3(t,x,y,\dot{x},\dot{y}) dt,$ $z_2 = \int
f_4(t,x,y,\dot{x},\dot{y}) dt$, $z_1\ne z_2$.  Now appropriately
restricting the
forms of $f_3$ and $f_4$, one can get six  different types of
linearizing transformations.
\begin{enumerate}
\item{ If $z_1$ and $z_2$
are perfect differential functions and $w_i$'s and $z_i$'s,
$i=1,2,$ do not contain the variables $\dot{x}$ and $\dot{y}$, then we call the resultant transformation, namely, $w_1=f_1(t,x,y)$,
$w_2=f_2(t,x,y)$, $z_1=f_3(t,x,y)$, $z_2=f_4(t,x,y)$, as a point
transformation of type-II.}
\item{ Suppose $z_1$ is a perfect differential
function and $z_2$ is not a perfect differential function or
vice versa, and if $z_1$
and $z_2$ do not contain the variables $\dot{x}$ and $\dot{y}$, then
we can call the resultant transformation, namely $w_1=f_1(t,x,y)$,
$w_2=f_2(t,x,y),\;z_1=f_3(t,x,y)$, $z_2=\int f_4(t,x,y)dt$ or
$z_1=\int f_3(t,x,y)dt$, $z_2=f_4(t,x,y)$, as a mixed
point-generalized Sundman transformation.}
\item{On the other hand, if any one of the independent variables
contains the variables $\dot{x}$ and $\dot{y}$, then we call the resultant
transformation, namely $w_1=f_1(t,x,y)$,
$w_2=f_2(t,x,y),\;z_1=f_3(t,x,y)$, $z_2=\int
f_4(t,x,y,\dot{x},\dot{y})dt$ or $z_1=\int
f_3(t,x,y,\dot{x},\dot{y})dt$ and  $z_2=f_4(t,x,y)$, as a
mixed point-generalized linearizing transformation.}
\item{ Suppose the independent variables
are not perfect differential functions and are also not functions of
$\dot{x}$ and $\dot{y}$, that is, $w_1=f_1(t,x,y)$,
$w_2=f_2(t,x,y),\;z_1=\int f_3(t,x,y)dt$, $z_2=\int f_4(t,x,y)dt$,
then we call the resultant transformation as a generalized Sundman
transformation of type-II.}
\item{ Further, if one of the independent variables, say $z_1$, does not contain the
derivative terms while the other independent variable $z_2$ does 
contain the derivative terms or
vice versa, that is
$w_1=f_1(t,x,y)$, $w_2=f_2(t,x,y)$, $z_1=\int f_3(t,x,y)dt$,
$z_2=\int f_4(t,x,y,\dot{x},\dot{y})dt$ or $z_1=\int
f_3(t,x,y,\dot{x},\dot{y})dt$, $z_2=\int f_4(t,$ $x,y)dt$, then we call
the resultant
transformation as a mixed generalized Sundman-generalized
linearizing transformation.}
\item{ As a general case, if we allow both the independent variables,
$z_1$ and $z_2$, to be non-perfect
differential functions and also to contain derivative terms,  that is,
$w_1=f_1(t,x,y)$,  $w_2=f_2(t,x,y)$, $z_1=\int
f_3(t,x,y,\dot{x},\dot{y})dt$, $z_2=\int
f_4(t,x,y,\dot{x},\dot{y})dt$, then the resultant
transformation will be termed as a generalized linearizing transformation of
type-II.}
\end{enumerate}
Finally the possibility that $w_1=f_1(t,x,y)$,
$w_2=f_2(t,x,y)$, $z_1=f_3(t,x,y,\dot{x},\dot{y})$ and
$z_2=f_4(t,x,y,\dot{x},\dot{y})$ is not considered in the present study and will be pursued separately.

\section{Applications}
In this section, we consider specific examples and illustrate each one of the
linearizing transformations identified in the previous section so as to make
clear the applicability of them under different situations.

\subsection{Class-A Linearizing Transformations ($z_1=z_2$)}
\subsubsection{Example:1 Point transformation of Type-I}
Let us consider the system of SNODEs,
\begin{eqnarray}
\ddot{x}+\frac{(\dot{x}y-\dot{y}x)^2}{2xy(x-y)}+\omega^2x=0,\quad \ddot{y}-\frac{(\dot{x}y-\dot{y}x)^2}{2xy(x-y)}+\omega^2y=0, \label{fcou10}
\end{eqnarray}
where $\omega$ is a arbitrary constant.  
The two first integrals associated with (\ref{fcou10}), which can be
obtained using the formulation given in \S2 of Part IV (Chandrasekar
\textit{et al.} 2009), can be written as
\begin{eqnarray}
I_1 &=&(\dot{x}+\dot{y})\sin(\omega t)-\omega(x+y)\cos(\omega t), \nonumber\\
I_2&=&\frac{1}{2\sqrt{xy}}((\dot{x}y+x\dot{y})\sin(\omega t)-2\omega xy\cos(\omega t)). \label{cou10}
\end{eqnarray}
Rewriting (\ref{cou10}) in the form (\ref{met13b})-(\ref{met13d}), we get
\begin{eqnarray}
I_1& =&\sin^2(\omega t)\frac {d}{dt}((x+y)cosec(\omega t))=\frac {dt}{dz_1}\frac {dw_1}{dt}
= \frac {dw_1}{dz_1},\nonumber\\
I_2& =&\sin^2(\omega t)\frac {d}{dt}(\sqrt{xy}cosec(\omega t))= \frac {dt}{dz_2}\frac {dw_1}{dt}=\frac {dw_2}{dz_2},\label{cou11b}
\end{eqnarray}
so that we can identify point transformation of type I as
\begin{eqnarray}
w_1=(x+y)\mbox{cosec}(\omega t),\quad w_2=\sqrt{xy}\mbox{cosec}(\omega t),\quad z_1=z_2=z=-\frac{1}{\omega}\cot(\omega t). \label{cou12}
\end{eqnarray}
Utilizing the transformation (\ref{cou12}), one can transform
(\ref{fcou10}) to a set of free
particle equations, namely $\frac{d^2w_1}{dz^2}=0$ and
$\frac{d^2w_2}{dz^2}=0$, so that
$w_1=I_1z+I_3$ and $w_2=I_2z+I_4$, where $I_3$ and $I_4$ are integration
constants. Substituting the expressions for $w_i$'s, $i=1,2,$ and $z$ in the
free particle solutions and rewriting the resultant expressions in terms of the old
variables $x$ and $y$, one obtains the general solution
for equation (\ref{fcou10}) in the form
\begin{eqnarray}
x(t)=-\frac{1}{2}(A\mp\sqrt{A^2-4B^2}),\quad
y(t)=\frac{1}{2}(A\pm\sqrt{A^2-4B^2}),\label{scou13}
\end{eqnarray}
where $A=I_1\cos(\omega t)+I_3\sin(\omega t)$ and $B=I_2\cos(\omega t)+I_4\sin(\omega t)$. Here we point out that the nonlinear system (\ref{fcou10}) admits amplitude independent frequency of oscillations. 

In the above example we have considered the new dependent variables $(w_1,w_2)$ and independent variable $z$ to be functions of $x,y$ and
$t$ only.  We will now consider examples which admit more general
transformations.

\subsubsection{Example:2 Generalized Sundman Transformation of Type-I}
Let us focus our attention on the two-dimensional Mathews-Lakshmanan
oscillator system of the form (Cari$\tilde{n}$ena \textit{et al.} 2004;
Chandrasekar \textit{et al.} 2009)
\begin{eqnarray}
&&\ddot{x}=\frac{\lambda(\dot{x}^2+\dot{y}^2
+\lambda(x\dot{y}-y\dot{x})^2)x-\alpha^2x}{(1+\lambda r^2)},\nonumber\\
&&\ddot{y}=\frac{\lambda(\dot{x}^2+\dot{y}^2
+\lambda(x\dot{y}-y\dot{x})^2)y-\alpha^2y}{(1+\lambda r^2)},
\label{ext01}
\end{eqnarray}
where $r=\sqrt{x^2+y^2}$ and $\lambda$ and $\alpha$ are arbitrary parameters.
For $\alpha=0$, equation (\ref{ext01}) admits the following two integrals of
motion, 
\begin{eqnarray}
\hat{I}_1 =\frac{(1+\lambda y^2)\dot{x}-\lambda xy\dot{y}}
{\sqrt{1+\lambda r^2}},\quad
\hat{I}_2=\frac{(1+\lambda x^2)\dot{y}-\lambda xy\dot{x}}{\sqrt{1+\lambda r^2}}.
\label{gfcou10a}
\end{eqnarray}
We note that the integrals $I_2$ and $I_3$ (vide equations (5.35)
and (5.36) in Part IV (Chandrasekar \textit{et al.} 2009)) can be
derived from (\ref{gfcou10a}) by using the relations
$I_2=-\lambda(\hat{I}_1^2+\hat{I}_2^2+\lambda I_1^2)$ and
$I_3=(\hat{I}_1+\hat{I}_2)^2-2\lambda I_1^2$, where $I_1$ is given
in (5.35) of Part IV. The general case ($\alpha\neq0$), can be linearized through mixed generalized Sundman - generalized linearized transformation (see example 8 below). To demonstrate the linearization through generalized Sundman transformation of Type-I, we here consider equation (\ref{ext01}) with $\alpha=0$. 

For the case $\alpha=0$, one may note that on making a substitution $y(t)=y(x(t))$ into (\ref{ext01}), one can obtain a non-autonomous second order ODE in $y(x)$. Though this equation satisfies the linearization condition for point transformation (Sookmee \& Meleshko 2008), finding the linearizing transformation and the general solution for the transformed ODE turns out to be nontrivial. On the other hand, we provide a straightforward procedure of linearization. 

The above two integrals (\ref{gfcou10a}) can be rewritten as
\begin{eqnarray}
\hat{I}_1& =\displaystyle{(1+\lambda r^2)\frac{d}{dt}
\bigg(\frac{x}{\sqrt{1+\lambda r^2}}\bigg)}=\frac {dt}{dz_1}\frac {dw_1}{dt}
= \frac {dw_1}{dz_1},\label{gfcou11a}\\
\hat{I}_2& =\displaystyle{(1+\lambda r^2)\frac{d}{dt}
\bigg(\frac{y}{\sqrt{1+\lambda r^2}}\bigg)}=\frac {dt}{dz_2}\frac {dw_2}{dt}=
 \frac {dw_2}{dz_2}.\label{gfcou11b}
\end{eqnarray}
From the above equation, we identify the new dependent and independent variables as
\begin{eqnarray}
w_1=\frac{x}{\sqrt{1+\lambda r^2}},\quad w_2=\frac{y}{\sqrt{1+\lambda r^2}},
\quad z_1=z_2=z=\int \frac{dt}{(1+\lambda r^2)}. \label{gfcou12}
\end{eqnarray}
One may observe that the independent variables $z_1$ and $z_2$ are
not perfect differentials even though they turn out to be
identical. By utilizing the above new variables, one can transform
equation (\ref{ext01}), with $\alpha=0$, to the free particle
equations, that is, $\frac{d^2w_1}{dz^2}=0$ and
$\frac{d^2w_2}{dz^2}=0$.

Unlike the earlier example, one cannot unambiguously integrate these two linear
equations in terms of the original variables due to the nonlocal nature of the independent variable. To overcome this
problem one should express $(1+\lambda r^2)$ in terms of either $z_1$ or $z_2$
so that the resultant expression $dz_1=dz_2=\frac{dt}{(1+\lambda r^2)}$ can be
integrated. In the following, we describe a procedure to obtain an expression for
the new independent variable.

Now integrating equation (\ref{gfcou11a}) and using first relation in (\ref{gfcou12}), we get
\begin{eqnarray}
\frac{x}{\sqrt{1+\lambda r^2}}&=&\hat{I}_1z_1,
\label {gfceq12a}
\end{eqnarray}
where we have fixed the integration constant to be zero (without loss of
generality). On the other hand from expressions (\ref{gfcou11a}) and (\ref{gfcou11b}) we get
$\frac {dw_1}{dw_2}=\frac {\hat{I}_1}{\hat{I}_2}$ from which one obtains
\begin{eqnarray}
w_1=\frac{\hat{I}_1}{\hat{I}_2}w_2+\hat{I}_3 \quad \Rightarrow \quad
\frac{x}{\sqrt{1+\lambda r^2}}
=\frac{\hat{I}_1}{\hat{I}_2}\frac{y}{\sqrt{1+\lambda r^2}}+\hat{I}_3,
\label {gfceq12b}
\end{eqnarray}
where $\hat{I}_3$ is the integration constant. Equation (\ref{gfceq12b})
provides us an identity
\begin{eqnarray}
\frac{y}{\sqrt{1+\lambda r^2}}=\hat{I}_2z_1-\frac{\hat{I}_2\hat{I}_3}{\hat{I}_1}. \label {gfceq12d}
\end{eqnarray}
Now squaring and adding the equations
(\ref{gfceq12a}) and (\ref{gfceq12d}), we obtain
\begin{eqnarray}
\frac{\lambda r^2}{1+\lambda r^2}=\lambda((\hat{I}_1^2+\hat{I}_2^2)z_1^2
-2\frac{\hat{I}_2^2\hat{I}_3}{\hat{I}_1}z_1
+\frac{\hat{I}_2^2\hat{I}_3^2}{\hat{I}_1^2}). \label {gfceq12e}
\end{eqnarray}
From (\ref{gfceq12e}), one can express $(1+\lambda r^2)$ in terms of $z_1$ as
\begin{eqnarray}
1+\lambda r^2=\frac{1}{1-\lambda\bigg((\hat{I}_1^2+\hat{I}_2^2)z_1^2
-\frac{2\hat{I}_2^2\hat{I}_3}{\hat{I}_1}z_1
+\frac{\hat{I}_2^2\hat{I}_3^2}{\hat{I}_1^2}\bigg)}. \label {gfceq12f}
\end{eqnarray}
Substituting (\ref{gfceq12f}) in the last relation given in equation
(\ref{gfcou12}), we arrive at the following integral relationship
between $z_1$ and $t$, namely,
\begin{eqnarray}
dz_1=\bigg(1-\lambda\bigg((\hat{I}_1^2+\hat{I}_2^2)z_1^2
-\frac{2\hat{I}_2^2\hat{I}_3}{\hat{I}_1}z_1
+\frac{\hat{I}_2^2\hat{I}_3^2}{\hat{I}_1^2}\bigg)\bigg)dt. \label {gfceq13}
\end{eqnarray}

Since the variables are separated now, one can integrate this equation and
obtain an expression which connects the new independent variable with the
old independent variable through the relation
\begin{eqnarray}
z_1=\frac{\sqrt{\lambda}\hat{I}_2^2\hat{I}_3-\hat{I}_1\omega\tan[\omega (t-t_0)]}
{\sqrt{\lambda}\hat{I}_1(\hat{I}_1^2+\hat{I}_2^2)},\qquad ({\hat{I}_2^2(\lambda \hat{I}_3^2-1)-\hat{I}_1^2})>0, \label {gfceq13a}
\end{eqnarray}
where $\omega=\sqrt{\lambda}\sqrt{{\hat{I}_2^2(\lambda \hat{I}_3^2-1)-\hat{I}_1^2}}$ and
$t_0$ is the fourth integration constant. From equations
(\ref{gfceq12a})-(\ref{gfceq12d}) and (\ref{gfceq12f}), we get
\begin{eqnarray}
&&x(t)=\hat{I}_1z_1\bigg[1-\lambda\bigg((\hat{I}_1^2+\hat{I}_2^2)z_1^2
-\frac{2\hat{I}_2^2\hat{I}_3}{\hat{I}_1}z_1
+\frac{\hat{I}_2^2\hat{I}_3^2}{\hat{I}_1^2}\bigg)\bigg]^{-\frac{1}{2}},
\nonumber\\
&&y(t)=\bigg(\hat{I}_2z_1-\frac{\hat{I}_2\hat{I}_3}{\hat{I}_1}\bigg)
\bigg[1-\lambda\bigg((\hat{I}_1^2+\hat{I}_2^2)z_1^2
-\frac{2\hat{I}_2^2\hat{I}_3}{\hat{I}_1}z_1
+\frac{\hat{I}_2^2\hat{I}_3^2}{\hat{I}_1^2}\bigg)\bigg]^{-\frac{1}{2}}.
\label {gfceq13b}
\end{eqnarray}

Substituting the expression (\ref{gfceq13a}) in the above equation
(\ref{gfceq13b})
and simplifying the resultant expressions, we arrive at the following general
solution for equation (\ref{ext01}) with $\alpha=0$ in the form
\begin{eqnarray}
&&x(t)=A\bigg(\lambda \hat{I}_2^2\hat{I}_3\cos[\omega (t-t_0)]
-\omega \hat{I}_1\sin[\omega(t-t_0)]\bigg),\nonumber\\
&&y(t)=-\hat{I}_2A
\bigg(\lambda \hat{I}_1\hat{I}_3\cos[\omega (t-t_0)]
+\omega \sin[\omega (t-t_0)]\bigg),\label {gfceq13c}
\end{eqnarray}
where $A=\frac{1}{\lambda(\hat{I}_1^2+\hat{I}_2^2)}
\sqrt{\frac{(\hat{I}_1^2+\hat{I}_2^2)}{\hat{I}_1^2
+\hat{I}_2^2(1-\lambda \hat{I}_3^2)}}$ and $\omega=\sqrt{\lambda}\sqrt{{\hat{I}_2^2(\lambda \hat{I}_3^2-1)-\hat{I}_1^2}}$.


\subsubsection{Example:3 Generalized linearizing transformation of Type-I }
In the previous example we restricted the new independent variable
to be a nonlocal one and be a function of $t$, $x$ and $y$ only. Now we
relax the latter condition and also allow the independent variables $z_1$
and $z_2$ to contain derivative terms, namely $\dot{x}$ and
$\dot{y}$. To illustrate this case let us consider a two coupled second order equations  of the form
\begin{eqnarray}
&&\ddot{x}=\frac{2\dot{x}\dot{y}(x\dot{x}+x^3)-2y\dot{x}^3}{x^3y},\nonumber\\
&&\ddot{y}=\frac{2\dot{x}\dot{y}(x^2y-y\dot{x}+x\dot{y})+x(x^2\dot{y}^2-y^2\dot{x}^2)}{x^3y}.
\label{lscou10}
\end{eqnarray}
One can easily identify two integrals for the equation (\ref{lscou10}) in the form
\begin{eqnarray}
I_1=y^2\bigg(\frac{1}{x^2}+\frac{1}{\dot{x}}\bigg) \quad \mbox{and} \quad
I_2 =y\bigg(\frac{y}{x}-\frac{\dot{y}}{\dot{x}}\bigg) .
\label{lscou11}
\end{eqnarray}
Rewriting the above integrals as
\begin{eqnarray}
I_1 =\frac{y^2}{\dot{x}}\frac {d}{dt}\bigg(t-\frac{1}{x}\bigg)= 
\frac {dw_1}{dz_1},\;\;
I_2 =\frac{y^2}{\dot{x}}\frac {d}{dt}\bigg(\log\bigg[\frac{x}{y}\bigg]\bigg)= \frac {dw_2}{dz_2},
\label{lscou11a}
\end{eqnarray}
we identify the following set of linearizing transformations for the equation
(\ref{lscou10}), that is
\begin{eqnarray}
w_1  = t-\frac{1}{x},\quad w_2=\log\bigg[\frac{x}{y}\bigg], \quad z_1  = 
z_2 =z=\int \frac{\dot{x}}{y^2} dt.
 \label{lscou12}
\end{eqnarray}
One may note that the independent variables are not only nonlocal
but also involve derivative terms. It is easy to check that
equation~(\ref{lscou12}) transforms (\ref{lscou10}) to the
linearized form (\ref{met13h}).

Again, as in the previous example, one cannot directly obtain the
solution for (\ref{lscou10}) by direct integration of the linear ODEs
due to the nonlocal nature of the independent variables. This can
be overcome by expressing the term $\dot{x}/y^2$ in terms of
either $z_1$ or $z_2$ so that the resultant equation can be
integrated to obtain an explicit form for  the new independent
variable in terms of the old variables as we discuss below.

From (\ref{lscou11}), we have
\begin{eqnarray}
\frac{\dot{x}}{y^2}=\frac{1}{(I_1-\frac{y^2}{x^2})}. \label {lsceq13a}
\end{eqnarray}
Since $\frac{dw_1}{dz_1}=I_1$ (from equation (\ref{lscou11a})), we have $w_1=I_1z_1$, so that
\begin{eqnarray}
t-\frac{1}{x}=I_1z_1, \label {lsceq13b}
\end{eqnarray}
where without loss of
generality we have fixed the integration constant to be zero.
On the other hand, from (\ref{lscou11a}), we have
$\frac{dw_1}{dw_2}=\frac{I_1}{I_2}$, which in turn gives
\begin{eqnarray}
w_1=\frac{I_1}{I_2}w_2+I_3. \label {lsceq13c}
\end{eqnarray}
In terms of old variables (\ref{lsceq13c}) reads as
\begin{eqnarray}
t-\frac{1}{x}=\frac{I_1}{I_2}\log\bigg[\frac{x}{y}\bigg]+I_3. \label {lsceq13d}
\end{eqnarray}
From the identites (\ref{lsceq13b}) and (\ref{lsceq13d}), we can express $y/x$
interms of $z_1$ in the form
\begin{eqnarray}
\frac{y}{x}=\exp[-\frac{I_2}{I_1}(I_1z_1-I_3)]. \label {lsceq13e}
\end{eqnarray}
Now substituting (\ref{lsceq13e}) into equation (\ref{lsceq13a}), we can express
$(\frac{\dot{x}}{y^2})$ in terms of $z_1$, and plugging the latter relation into
the third relation in (\ref{lscou12}), we arrive at
\begin{eqnarray}
dt=(I_1-\exp[-2\frac{I_2}{I_1}(I_1z_1-I_3)])dz_1.
\label {lsceq14}
\end{eqnarray}
Integrating the above equation, we obtain
\begin{eqnarray}
t+t_0=I_1z_1+\frac{e^{-\frac{2I_2}{I_1}(I_1z_1-I_3)}}{2I_2},\label {lsceq15}
\end{eqnarray}
where $t_0$ is the fourth integration constant. Substituting the 
expression $z_1=\frac{t-\frac{1}{x}}{I_1}$ (vide equation (\ref{lsceq13b})) into equations (\ref{lsceq13e}) and (\ref{lsceq15}), we obtain the general solution of (\ref{lscou10}) in the implicit form
\begin{eqnarray}
x(t)=-\frac{1}{t_0}+\frac{x(t)e^{-\frac{2I_2}{I_1}(t-\frac{1}{x(t)}-I_3)}}{2I_2t_0},\quad y(t)=x(t)\exp[-\frac{I_2}{I_1}(t-\frac{1}{x(t)}-I_3)].
\label {lsceq16}
\end{eqnarray}

\subsection{Class-B Linearizing Transformations ($z_1\neq z_2$)}
In the Class-A category, in all the three examples, we considered that the new
independent variables, $z_1$ and $z_2$, are identical. However, this need
not always be the case in the theory of linearizing transfomations,
as discussed in Sec. \S2. We now present specific examples to illustrate more general transformations.
\subsubsection{Example:4 Point transformation of Type-II}
Let us consider a quasi periodic oscillator governed by a set of two coupled SNODEs of the form
\begin{eqnarray}
\ddot{x}+\frac{(\dot{x}y-\dot{y}x)^2+2x^2y(\omega_1^2(x+y)-2\omega_2^2y)}{2xy(x-y)}=0,\nonumber\\ \ddot{y}-\frac{(\dot{x}y-\dot{y}x)^2+2xy^2(\omega_1^2(x+y)-2\omega_2^2x)}{2xy(x-y)}=0.
 \label{tcou10}
\end{eqnarray}

To explore the linearizing transformation for equation (\ref{tcou10}), we consider the two associated integrals,
\begin{eqnarray}
I_1&=&(\dot{x}+\dot{y})\cos{\omega_1t}+\omega_1(x+y)\sin{\omega_1 t},\nonumber\\
I_2&=&\frac{1}{2\sqrt{xy}}((\dot{x}y+\dot{y}x)\sin{\omega_2t}-2\omega_2xy\cos{\omega_2t}).
\label{tcou10a}
\end{eqnarray}
Rewriting (\ref{tcou10a}) in the form
\begin{eqnarray}
I_1& =&\cos^2(\omega_1 t)\frac {d}{dt}((x+y)\sec(\omega_1 t))= \frac {dw_1}{dz_1},\label{tcou11a}\\
I_2& =&\sin^2(\omega_2 t)\frac {d}{dt}(\sqrt{xy}\mbox{cosec}(\omega_2 t))= \frac
{dw_2}{dz_2},\label{tcou11b}
\end{eqnarray}
one can identify the new dependent and the independent variables as
\begin{eqnarray}
w_1  = (x+y)\sec(\omega_1 t),\; z_1  =\frac{1}{\omega_1}\tan(\omega_1 t),\nonumber\\ 
w_2  =\sqrt{xy}\mbox{cosec}(\omega_2 t), \; z_2  =-\frac{1}{\omega_1}\cot(\omega_2 t).
\label{tcou12}
\end{eqnarray}
One may note that now the independent variables $z_1$ and $z_2$ are not the same.

The new variables transform (\ref{tcou10}) to the free particle
equations $\frac{d^2w_1}{dz_1^2}=0$ and $\frac{d^2w_2}{dz_2^2}=0$.
From the general solutions
$w_1=I_1z_1+I_3$ and $w_2=I_2z_2+I_4$, where $I_i$'s, $i=1,2,3,4,$ are the
integration constants, and using the  expressions for $w_i$'s and $z_i$'s,
$i=1,2$, given in (\ref{tcou12}), we arrive at the general solution for the equation (\ref{tcou10}) in the form
\begin{eqnarray}
x(t)=-\frac{1}{2}(A\mp\sqrt{A^2-4B^2}),\quad
y(t)=\frac{1}{2}(A\pm\sqrt{A^2-4B^2}),\label{tcou13}
\end{eqnarray}
where now $A=I_1\sin(\omega_1 t)+I_3\cos(\omega_1 t)$ and $B=I_2\cos(\omega_2 t)+I_4\sin(\omega_2 t)$.

\subsubsection{Example:5 Mixed Point-Generalized Sundman transformation (PGST)}
Let us consider the two dimensional force-free coupled Duffing-van der Pol oscillator (DVP) equation of the form
\begin{eqnarray}
&&\ddot{x}+4(\alpha+\beta(k_1x+k_2y)^2)\dot{x}+\alpha(3\alpha+4\beta(k_1x+k_2y)^2)x=0,
\nonumber\\
&&\ddot{y}+4(\alpha+\beta(k_1x+k_2y)^2)\dot{y}+\alpha(3\alpha+4\beta(k_1x+k_2y)^2)y=0. \label{gscou10}
\end{eqnarray}

One may note that the point transformation $X=k_1x+k_2y$ and $Y=k_1x-k_2y$ helps one to rewrite (\ref{gscou10}) in a separable form,
\begin{subequations}
\begin{eqnarray}
&&\ddot{X}+4(\alpha+\beta X^2)\dot{X}+\alpha(3\alpha+4\beta X^2)X=0,\label{gscou10aa}\\
&&\ddot{Y}+4(\alpha+\beta X^2)\dot{Y}+\alpha(3\alpha+4\beta X^2)Y=0. \label{gscou10bb}
\end{eqnarray}
\end{subequations}
The solution to equation (\ref{gscou10aa}) can only be obtained in implicit form (Chandrasekar \textit{et al.} 2005). Consequently, equation (\ref{gscou10bb}) cannot be solved explicitly in this way. Further, the linearization of the scalar DVP oscillator (\ref{gscou10aa}) itself has not yet been reported. In the following we use our procedure to find the linearizing transformation and general solution to (\ref{gscou10}) straightforwardly.

The first two integrals for (\ref{gscou10}) can be easily identified using the procedure given in Chandrasekar \textit{et al.} (2009) in
the form
\begin{eqnarray}
I_1&=&\bigg(\frac{x\dot{y}-y\dot{x}}{\dot{y}+\alpha y}\bigg)
e^{\alpha t},\nonumber\\
I_2& =&(k_1\dot{x}+k_2\dot{y}+\alpha(k_1x+k_2y)+\frac{4\beta}{3}(k_1x+k_2y)^3)
e^{3\alpha t}.
\end{eqnarray}
The above integrals can be rewritten as
\begin{eqnarray}
I_1&=&-\frac{y^2e^{\alpha t}}{(\dot{y}+\alpha y)}\frac {d}{dt}(\frac{x}{y})
= \frac {dw_1}{dz_1},\label{gscou11}\\
I_2 &=&-((k_1x+k_2y)e^{\frac{5}{3}\alpha t})^3\frac {d}{dt}\bigg[\bigg(\frac{1}{2}(k_1x+k_2y)^{-2}+\frac{2\beta}{3\alpha}\bigg)e^{-2\alpha t}\bigg]=
 \frac {dw_1}{dz_2},\nonumber
\end{eqnarray}
from which we can obtain the following linearizing transformations,
\begin{eqnarray}
w_1&=&\frac{x}{y},\quad \quad w_2=(\frac{1}{2}(k_1x+k_2y)^{-2}+\frac{2\beta}{3\alpha})e^{-2\alpha t},\nonumber\\
z_1&=&\frac{e^{-\alpha t}}{y},\quad z_2=-\int ((k_1x+k_2y)e^{\frac{5}{3}\alpha t})^{-3} dt. \label{gscou12}
\end{eqnarray}
One may note that in the present problem one of the new independent variables,
that is, $z_2$ is in a nonlocal form. In terms of the above new variables, equation
(\ref{gscou10}) assumes the forms $\frac{d^2w_1}{dz_1^2}=0$ and $\frac{d^2w_2}{dz_2^2}=0$.

Now we seek the general solution of (\ref{gscou10}) from the linearized
equations. Integrating
$\frac{d^2w_1}{dz_1^2}=0$, we obtain
\begin{eqnarray}
w_1=I_1z_1+I_3, \label {gsceq12a}
\end{eqnarray}
where $I_3$ is the integration constant. Rewriting (\ref{gsceq12a}) in terms of the
old variables, we get
\begin{eqnarray}
x=I_1e^{-\alpha t}+I_3y.
\label {gsceq12cc}
\end{eqnarray}
However, the second linear equation,
$\frac{d^2w_2}{dz_2^2}=0$, cannot be integrated straightforwardly (in terms of
the original variables) due to the
nonlocal nature of the second independent variable. To obtain an explicit
form for $z_2$, we rewrite $I_2$ in the
integral form (vide equation (\ref{gscou11})) to obtain
\begin{eqnarray}
(\frac{1}{2}(k_1x+k_2y)^{-2}+\frac{2\beta}{3\alpha})e^{-2\alpha t}=-I_2\int [(k_1x+k_2y)e^{\frac{5}{3}\alpha t}]^{-3} dt =I_2z_2. \label {gsceq12b}
\end{eqnarray}
Equation (\ref{gsceq12b}) provides us a relation
\begin{eqnarray}
(k_1x+k_2y)^{-1}=\sqrt{2I_2z_2e^{2\alpha t}-\frac{4\beta}{3\alpha}}. \label {gsceq12c}
\end{eqnarray}
Now substituting the relation (\ref{gsceq12c}) into the non-local variable $z_2$
(vide equation (\ref{gscou12})), one gets
\begin{eqnarray}
\frac{dz_2}{dt}=-I_2 (2I_2z_2-\frac{4\beta}{3\alpha}e^{-2\alpha t})^{\frac{3}{2}}
e^{-2\alpha t}.
 \label {gsceq12e}
\end{eqnarray}
Solving the above equation, we obtain
\begin{eqnarray}
&&t_0-\frac{1}{2\alpha}e^{2\alpha t}=\frac {a}{3I_2}\bigg[\sqrt{3}\,\tan^{-1}\left(\frac{\sqrt{3}}{2a(2I_2z_2-\frac{4\beta}{3\alpha}e^{-2\alpha t})^{\frac{1}{2}}-1}\right)\label {gsceq12f}\\
&&\qquad\qquad
+\frac{1}{2}
\log\bigg(\frac{(1+a(2I_2z_2-\frac{4\beta}{3\alpha}e^{-2\alpha t})^{\frac{1}{2}})^2}{1-a(2I_2z_2-\frac{4\beta}{3\alpha}e^{-2\alpha t})^{\frac{1}{2}}+a^2(2I_2z_2e^{2\alpha t}-\frac{4\beta}{3\alpha}e^{-2\alpha t})}\bigg)
\bigg], \nonumber
\end{eqnarray}
where $a=\sqrt[3]{-\frac {3I_2}{4\beta}}$ and $t_0$ is the
fourth integration constant. From the expression (\ref{gsceq12f}) and equations (\ref{gsceq12cc})
and (\ref{gsceq12c}) one can deduce the general solution for (\ref{gscou10}) in implicit form. The resultant expression coincides exactly with equation (6.18) given in Chandrasekar \textit{et al.} (2009).

In the present example we considered one of the independent variables to be in a
nonlocal form. As we have two independent variables one can also have the possibility
of having both the independent variables to be of nonlocal form. Indeed this is the case of our next example.

\subsubsection{Example:6 Generalized Sundman transformation of Type-II (GST-II)}
To illustrate the GST-II, we consider the equation of the form
\begin{eqnarray}
\ddot{x}-\frac{2}{(x^2+y^2)}((\dot{x}^2-\dot{y}^2)x+2y\dot{x}\dot{y})+\frac{2}{t^2}x=0,
\nonumber\\
\ddot{y}-\frac{2}{(x^2+y^2)}(2x\dot{x}\dot{y}-(\dot{x}^2-\dot{y}^2)y)+\frac{2}{t^2}y=0. \label{gs2cou10}
\end{eqnarray}
The first two integrals for (\ref{gs2cou10}) can be evaluated as 
\begin{eqnarray}
I_1 &=&\frac{(2(t\dot{y}-y)xy)+(x^2-y^2)(t\dot{x}-x)}{t^2(x^2+y^2)^2},
\nonumber\\
I_2 &=&\frac{(2xy(t\dot{x}-x)+(x^2-y^2)(t\dot{y}-y))}{t^2(x^2+y^2)^2}.
\label{gt2cou11}
\end{eqnarray}
Rewriting these two integrals as
\begin{eqnarray}
I_1 &=&\frac{x^2}{(x^2+y^2)^2}\frac {d}{dt}\bigg(\frac{x}{t}+\frac{y^2}{tx}\bigg)= \frac {dw_1}{dz_1},\nonumber\\
I_2 &=&\frac{y^2}{(x^2+y^2)^2}\frac {d}{dt}\bigg(\frac{y}{t}+\frac{x^2}{ty}\bigg)= \frac {dw_2}{dz_2},
\label{gtcou11}
\end{eqnarray}
we identify the linearizing transformations in a more general form,
\begin{eqnarray}
w_1=\frac{x}{t}+\frac{y^2}{tx},\; w_2=\frac{y}{t}+\frac{x^2}{ty},\; z_1=\int \frac{(x^2+y^2)^2}{x^2} dt,\; z_2=\int \frac{(x^2+y^2)^2}{y^2} dt.
\label{gtcou12}
\end{eqnarray}
The GST-II, (\ref{gtcou12}), takes equation (\ref{gscou10}) to
the free particle equations, $\frac{d^2w_1}{dz_1^2}=0$ and
$\frac{d^2w_2}{dz_2^2}=0$. To obtain the solution in
terms of the original variables, we have to replace both $\int ((x^2+y^2)^2/x^2) dt$ and $\int ((x^2+y^2)^2/y^2) dt$ by the variables $z_1$ and $t$, and $z_2$ and $t$, respectively, and integrate the resultant equations.

To do so, first we rewrite the first integrals $I_1$ and $I_2$ given
by (\ref{gtcou11}) in integral forms,
which in turn lead us to $w_1=I_1z_1$
and $w_2=I_2z_2$. Since $w_1$ and $w_2$ do not contain nonlocal variables
we can replace them by the old variables (vide equation
(\ref{gtcou12})), that is
\begin{eqnarray}
\frac{x}{t}+\frac{y^2}{tx}=I_1z_1,\qquad \frac{y}{t}+\frac{x^2}{ty}=I_2z_2, \label {gtceq13}
\end{eqnarray}
where we have fixed the integration constants to be zero (without loss of
generality).

We observe that to integrate the last two expressions in (\ref{gtcou12}) one
should further replace $z_1$ and $z_2$ in terms of $t$. So we substitute the
above expressions for $x$ and $y$ in terms of $z_1$ and $z_2$, respectively, in the
last two relations in (\ref{gtcou12}), and obtain
\begin{eqnarray}
dz_1=I_1^2z_1^2t^2dt \quad \mbox{and}\quad dz_2=I_2^2z_2^2t^2 dt.
\label {gtceq14}
\end{eqnarray}
Now integrating both the equations, we get
\begin{eqnarray}
z_1=(I_1^2(I_3-\frac{t^3}{3}))^{-1}\quad \mbox{and}\quad  z_2=(I_2^2(I_4-\frac{t^3}{3}))^{-1},
\label {gtceq14a}
\end{eqnarray}
where $I_3$ and $I_4$ are the third and forth integration constants, respectively. 
Plugging equation (\ref{gtceq14a}) into
(\ref{gtceq13}), we arrive at the following general solution for the
equation (\ref{gs2cou10}),
\begin{eqnarray}
x(t)=\frac{3t(I_1(3\hat{I}_3-(I_1^2-I_2^2)t^3)-I_2(3\hat{I}_4+2I_1I_2t^3)}{(3\hat{I}_3-(I_1^2-I_2^2)t^3)^2+(3\hat{I}_4+2I_1I_2t^3)^2},\nonumber\\
y(t)=\frac{3t(I_2((I_1^2-I_2^2)t^3-3\hat{I}_3)-I_1(3\hat{I}_4+2I_1I_2t^3)}{(3\hat{I}_3-(I_1^2-I_2^2)t^3)^2+(3\hat{I}_4+2I_1I_2t^3)^2},\label {gtceq13ab}
\end{eqnarray}
where $\hat{I}_3=(I_1I_3-I_2I_4)/(I_1^3+I_1I_2^2)$ and $\hat{I}_4=(I_2I_3+I_1I_4)/(I_2^3-I_1^2I_2)$.

In the previous two examples we focussed our attention on the case in which the new
independent variable(s) is(are) nonlocal and does(do) not admit velocity
dependent terms. Now we relax this condition and allow either one or both the
independent variables to admit velocity dependent terms but in nonlocal form.

\subsubsection{Example:7 Mixed Point-Generalized linearizing transformation}
To demonstrate this, we consider a
variant of the two-dimensional Mathews and Lakshmanan equation (\ref{ext01}) of the form
\begin{eqnarray}
\ddot{x}=\frac{\lambda(\dot{x}^2+\dot{y}^2
+2 \lambda(\dot{y}-\dot{x})^2)-\alpha^2}{(1+2\lambda (x+y))},\quad
\ddot{y}=\frac{\lambda(\dot{x}^2+\dot{y}^2
+2 \lambda(\dot{y}-\dot{x})^2)-\alpha^2}{(1+2\lambda (x+y))}.
\label{fglt01}
\end{eqnarray}
Equation (\ref{fglt01}) admits the following two integrals of motion,
\begin{eqnarray}
I_1 =\dot{x}-\dot{y},\quad
I_2=\frac{\alpha^2-\lambda((1+2\lambda)(\dot{y}-\dot{x})^2+2\dot{x}\dot{y})}
{(1+2\lambda (x+y))}.
\label{fglt02}
\end{eqnarray}
Rewriting (\ref{fglt02}) in the form
\begin{eqnarray}
I_1& =&\displaystyle{\frac{d}{dt}(x-y)},\label{fglt03a}\\
I_2& =&\displaystyle{\frac{\alpha^2
-\lambda((1+2\lambda)(\dot{y}-\dot{x})^2+2\dot{x}\dot{y})}{2\lambda (\dot{x}+\dot{y})}
\frac{d}{dt}\bigg[\log[1+2\lambda (x+y)]\bigg]},\label{fglt03b}
\end{eqnarray}
one can easily identify the linearizing transformations for (\ref{fglt01}) as
\begin{eqnarray}
&w_1=(x-y),\quad &w_2=\log[1+2\lambda (x+y)],\nonumber\\
&z_1=t,\quad &z_2=\int\frac{2\lambda (\dot{x}+\dot{y})}
{\alpha^2-\lambda((1+2\lambda)(\dot{y}-\dot{x})^2+2\dot{x}\dot{y})}dt.
\label{fglt04}
\end{eqnarray}
In terms of the above new variables, equation (\ref{fglt01}) gets transformed to
the free
particle equations (\ref{met13h}). One may note that one of the new
independent variables is not only in non-local form but also contains derivative
terms which in turn complicates the situation to obtain the general solution.

Since $w_1$ and $z_1$ are both of point transformation types, one can integrate
the first free particle equation, namely, $\frac{d^2w_1}{dz_1^2}=0$ and obtain
$w_1=I_1z_1+I_3$, where $I_3$ is an integration constant. Replacing the latter in terms of the old variables, one gets the relation
$(x-y)=I_1t+I_3$. On the other hand, from the solution of the second linear equation
$\frac{d^2w_2}{dz_2^2}=0$, we can write $w_2=I_2z_2\Rightarrow
\log[1+2\lambda (x+y)]=I_2z_2$ (again we assume the integration constant to be zero without loss of generality).

To evaluate $z_2$, let us first substitute (\ref{fglt02}) into (\ref{fglt04}), and
rewrite the latter in the form
\begin{eqnarray}
dz_2=\frac{2\lambda (\dot{x}+\dot{y})}
{I_2(1+2\lambda (x+y))}dt=\frac{2\lambda (I_1+2\dot{y})}
{I_2(1+2\lambda (x+y))}dt. \label {fglt05}
\end{eqnarray}
Now subsituting the form of $\dot{y}$ (vide equation (\ref{fglt02})), that is,
\begin{eqnarray}
\dot{y}=\frac{1}{2\lambda}
\bigg(-\lambda I_1\pm
\sqrt{2\lambda\alpha^2-\lambda^2(1+4\lambda)I_1^2
-2\lambda I_2(1+2\lambda (x+y))}\bigg), \label {fglt06}
\end{eqnarray}
into equation (\ref{fglt05}), and using the relation
$(1+2\lambda (x+y))=e^{I_2z_2}$, we obtain
\begin{eqnarray}
dz_2=\frac{2\sqrt{\lambda}\sqrt{2\alpha^2-\lambda(1+4\lambda)I_1^2-2I_2e^{I_2z_2}}}
{I_2e^{I_2z_2}}dt. \label {fglt07}
\end{eqnarray}
Integrating (\ref{fglt07}), we get
\begin{eqnarray}
z_2=\frac{1}{I_2}\log\bigg(\frac{2\alpha^2-\lambda(1+4\lambda)I_1^2
-4I_2^2\lambda (t-t_0)^2}{2I_2}\bigg),\label {fglt08}
\end{eqnarray}
where $t_0$ is an integration constant. Substituting the expression
(\ref{fglt08}) into the relation $2\lambda (x+y)=e^{I_2z_2}-1$, we get
\begin{eqnarray}
x+y=\frac{2\alpha^2-\lambda(1+4\lambda)I_1^2-4I_2^2\lambda (t-t_0)^2-2I_2}
{4\lambda I_2}.\label {fglt09}
\end{eqnarray}
From equation (\ref{fglt09}), and the relation $(x-y)=I_1t+I_3$, we
obtain the general solution for (\ref{fglt01}) in the form
\begin{eqnarray}
&&x(t)=\frac{2\alpha^2-\lambda(1+4\lambda)I_1^2-4I_2^2\lambda
(t-t_0)^2-I_2(2-4\lambda (I_1t+I_3))}
{8\lambda I_2},\nonumber\\
&&y(t)=\frac{2\alpha^2-\lambda(1+4\lambda)I_1^2-4I_2^2\lambda
(t-t_0)^2-I_2(2+4\lambda (I_1t+I_3))}
{8\lambda I_2}.\label {fglt10}
\end{eqnarray}
In this example we considered the case in which only one of the independent variables
is in nonlocal form. Now we consider the case in which both the independent
variables are in nonlocal forms.

\subsubsection{Example:8 Mixed Generalized Sundman-Generalized linearizing
transformation}
To illustrate this type of linearizing transformation let us consider again
equation (\ref{ext01}), but now with $\alpha\neq0$. Equation (\ref{ext01}) admits 
the following two integrals of motion, 
\begin{align}
I_1 =(y\dot{x}-x\dot{y}), \qquad
I_2=\frac{(\alpha^2-\lambda(\dot{x}^2+\dot{y}^2+\lambda(y\dot{x}-x\dot{y})^2))}
{1+\lambda r^2}.\label{ext04}
\end{align}
Rewriting these two integrals in the form
\begin{eqnarray}
\hat{I}_1& =&\displaystyle{y^2\frac{d}{dt}
\bigg(\frac{x}{y}\bigg)}= \frac {dw_1}{dz_1},\label{sgcou11a}\\
\hat{I}_2& =&\displaystyle{\frac{(\alpha^2-\lambda(\dot{x}^2+\dot{y}^2+\lambda(y\dot{x}-x\dot{y})^2)}{2\lambda(x\dot{x}+y\dot{y})}
\frac{d}{dt}
\bigg(\log(1+\lambda r^2)\bigg)}= \frac {dw_2}{dz_2},\label{sgcou11b}
\end{eqnarray}
and identifying the linearizing transformations, we obtain
\begin{eqnarray}
w_1&=&\frac{x}{y},\quad w_2=\log(1+\lambda r^2),\quad z_1=\int \frac{dt}{y^2},\nonumber\\
 z_2&=&\int
\frac{2\lambda(x\dot{x}+y\dot{y})}{(\alpha^2-\lambda(\dot{x}^2+\dot{y}^2+\lambda(y\dot{x}-x\dot{y})^2)} dt.
\label{sgcou12}
\end{eqnarray}
One can check that equation~(\ref{sgcou12}) transforms (\ref{ext01})
to the form (\ref{met13h}).

Rewriting the first integrals $I_1$ and $I_2$ in the integral form and
identifying them in terms of the new variables, we have $w_1=I_1z_1$ and
$w_2=I_2z_2$, which in turn also gives us a relationship between $x$ and $y$ with $z_1$ and
$z_2$, respectively (after fixing the integration constants to be zero without loss of generality), that is,
\begin{eqnarray}
x=I_1z_1y,\qquad 1+\lambda r^2=e^{I_2z_2}. \label {sgceq13}
\end{eqnarray}
Expressing $\dot{x}$ and $\dot{y}$ in terms of $I_1,I_2,x$ and $y$ (by utilizing the equation (\ref{ext04})) and substituting them in the expression for $dz_2$, we obtain
\begin{eqnarray}
dz_2=2\frac{(\lambda\alpha^2 r^2-\lambda(1+\lambda r^2)(\lambda I_1^2+I_2r^2)^{\frac{1}{2}}}{I_2(1+\lambda r^2)}dt.
\label {sgceq14a}
\end{eqnarray}
Now, from the expression for $1+\lambda r^2$ (vide equation (\ref{sgceq13})), we obtain 
\begin{eqnarray}
dz_2=\frac{2}{I_2}\bigg[(\alpha^2-\lambda^2I_1^2+I_2)e^{-I_2z_2}-I_2-\alpha^2e^{-2I_2z_2}\bigg]^{\frac{1}{2}}dt.
\label {sgceq14b}
\end{eqnarray}
Integrating the above equation we get
\begin{eqnarray}
I_3-t=\frac{1}{2\sqrt{I_2}}\tan^{-1}\bigg[\frac{-2I_2+(\alpha^2-\lambda^2I_1^2+I_2)e^{-I_2z_2}}
{2\sqrt{I_2}((\alpha^2-\lambda^2I_1^2+I_2)e^{-I_2z_2}-I_2-\alpha^2e^{-2I_2z_2})^{\frac{1}{2}}}\bigg], \label {sgceq14}
\end{eqnarray}
where $I_3$ is an integration constant which is nothing but the third integral of motion. 
In order to find the fourth integral, using $dz_1= \frac{dt}{y^2}$ and equation (\ref{sgceq14b}), we eliminate $dt$ to get
\begin{eqnarray}
dz_1=\frac{I_2}{2}\bigg[\frac{dz_2}
{y^2((\alpha^2-\lambda^2I_1^2+I_2)e^{-I_2z_2}-I_2-\alpha^2e^{-2I_2z_2})^{\frac{1}{2}}}\bigg]. \label {sgceq14c}
\end{eqnarray}
From equation (\ref{sgceq13}), we get $y^2=(e^{I_2z_2}-1)/(\lambda(I_1^2z_1^2+1))$, which on substitution into equation (\ref{sgceq14c}) leads to 
\begin{eqnarray}
\frac{dz_1}{(I_1^2z_1^2+1)}=\frac{\lambda I_2}{2}\bigg[\frac{(e^{I_2z_2}-1)^{-1}dz_2}
{y^2((\alpha^2-\lambda^2I_1^2+I_2)e^{-I_2z_2}-I_2-\alpha^2e^{-2I_2z_2})^{\frac{1}{2}}}\bigg], \label {sgceq14d}
\end{eqnarray}
Now integrating (\ref{sgceq14d}) we get
\begin{eqnarray}
I_4=\tan^{-1}\bigg[I_1z_1\bigg]-\frac{1}{2}\tan^{-1}\bigg[\frac{(I_2-\alpha^2-\lambda^2I_1^2)
+(\alpha^2-\lambda^2I_1^2-I_2)e^{I_2z_2}}
{2\lambda I_1((\alpha^2-\lambda^2I_1^2+I_2)e^{I_2z_2}-I_2e^{2I_2z_2}-\alpha^2)^{\frac{1}{2}}}\bigg], \label {sgceq14e}
\end{eqnarray}
where $I_4$ is the fourth integration constant. Now making use of these four integrals of motion, namely equations (\ref{ext04}), (\ref{sgceq14}) and (\ref{sgceq14e}), the general solution to equation (\ref{ext01}) can be straightforwardly constructed. The resultant solution also agrees with the equation (5.40) of Chandrasekar \textit{et al.} (2009), obtained through the modified PS approach, after a redefinition of integration constants.

\subsubsection{Example:9 Generalized linearizing transformation of Type-II}
To understand the generalized linearizing transformation let us start
with the following system of coupled second order ODEs,
\begin{eqnarray}
\ddot{x}+\frac{k(x\dot{x}-y\dot{y})+k^2x}{(x^2-y^2)}+\lambda x=0,\;\;
\ddot{y}+\frac{k(x\dot{y}-y\dot{x})-k^2y}{(x^2-y^2)}+\lambda y=0. \label{case903}
\end{eqnarray}
The associated first integrals are
\begin{eqnarray}
I_1=\bigg(\frac{k+\sqrt{-\lambda}(x+y)+\dot{x}+\dot{y}}{k-\sqrt{-\lambda}(x+y)+\dot{x}+\dot{y}}\bigg)e^{-2\sqrt{-\lambda}t},\nonumber\\
I_2 =\bigg(\frac{k+\sqrt{-\lambda}(x-y)+\dot{x}-\dot{y}}{k-\sqrt{-\lambda}(x-y)+\dot{x}-\dot{y}}\bigg)e^{-2\sqrt{-\lambda}t}.
\label{case905}
\end{eqnarray}
Rewriting (\ref{case905}) in the form
\begin{eqnarray}
I_1 &=&\frac{ke^{-3\sqrt{-\lambda}t}}{(k-\sqrt{-\lambda}(x+y)+\dot{x}+\dot{y})}
\frac{d}{dt}\bigg[\bigg(\frac{1}{\sqrt{-\lambda}}+\frac{x+y}{k}\bigg)e^{\sqrt{-\lambda}t}\bigg], \nonumber\\
I_2 &=&\frac{ke^{-3\sqrt{-\lambda}t}}{(k-\sqrt{-\lambda_4}(x-y)+\dot{x}-\dot{y})}
\frac{d}{dt}\bigg[\bigg(\frac{1}{\sqrt{-\lambda}}+\frac{x-y}{k}\bigg)e^{\sqrt{-\lambda}t}\bigg],
\end{eqnarray}
and identifying the new variables, we get the linearizing transformation,
\begin{eqnarray}
&&w_1  = (\frac{1}{\sqrt{-\lambda}}+\frac{x+y}{k})e^{\sqrt{-\lambda}t},
\;z_1  = \int \frac{(1-\frac{\sqrt{-\lambda}}{k}(x+y)+
\frac{\dot{x}+\dot{y}}{k})}{e^{3\sqrt{-\lambda}t}}dt,\nonumber\\
&&w_2=(\frac{1}{\sqrt{-\lambda}}+\frac{x-y}{k})e^{\sqrt{-\lambda}t},\;
z_2=\int \frac{(1-\frac{\sqrt{-\lambda}}{k}(x-y)+
\frac{\dot{x}-\dot{y}}{k})}{e^{3\sqrt{-\lambda}t}}dt.
 \label{case906}
\end{eqnarray}

From the first integrals we have (after assuming the integration constants to be zero without loss of generality)
\begin{eqnarray}
w_1=I_1z_1,\qquad w_2=I_2z_2. \label {case907}
\end{eqnarray}
Utilizing (\ref{case906}) in (\ref{case907}), we get
\begin{eqnarray}
x(t)=\frac{k}{2}(I_1z_1+I_2z_2)e^{-\sqrt{-\lambda}t}-\frac{k}{\sqrt{-\lambda}},\qquad 
y(t)=\frac{k}{2}(I_1z_1-I_2z_2)e^{-\sqrt{-\lambda}t}. 
\label {case908}
\end{eqnarray}
Substituting the expressions of $x$ and $y$ into equation (\ref{case905}), and 
solving the resultant equation for $\dot{x}$ and $\dot{y}$, we get
\begin{eqnarray}
\dot{x} & = &-k-\sqrt{-\lambda}\bigg[\frac{k}{2}(I_1z_1+I_2z_2)e^{-\sqrt{-\lambda}t}-\frac{k}{\sqrt{-\lambda}}\bigg]\bigg(\frac{1+e^{-\sqrt{-\lambda}t}}{1-e^{-\sqrt{-\lambda}t}}\bigg)\nonumber\\
\dot{y} & =& -\sqrt{-\lambda}\bigg[\frac{k}{2}(I_1z_1-I_2z_2)e^{-\sqrt{-\lambda}t}\bigg]\bigg(\frac{1+e^{-\sqrt{-\lambda}t}}{1-e^{-\sqrt{-\lambda}t}}\bigg).
 \label{case910}
\end{eqnarray}
Substituting the equations (\ref{case908}) and (\ref{case910}) into the expressions (\ref{case906}) for $dz_1$ and $dz_2$, and integrating the resultant equation, we obtain
\begin{eqnarray}
z_1 & =& \frac{1}{I_1}\bigg(\frac{e^{\sqrt{-\lambda}t}}{\sqrt{-\lambda}}+\bigg((e^{2\sqrt{-\lambda}t}-I_1)(I_3+\tanh^{-1}[\frac{e^{\sqrt{-\lambda}t}}{\sqrt{I_1}}])\bigg)\bigg)\nonumber\\
z_2 & =& \frac{1}{I_2}\bigg(\frac{e^{\sqrt{-\lambda}t}}{\sqrt{-\lambda}}+\bigg((e^{2\sqrt{-\lambda}t}-I_2)(I_4+\tanh^{-1}[\frac{e^{\sqrt{-\lambda}t}}{\sqrt{I_2}}])\bigg)\bigg),
 \label{case911}
\end{eqnarray}
where $I_3$ and $I_4$ are the third and fourth integration constants, respectively. From  (\ref{case911}) and (\ref{case908}), we can obtain the general solution for
equation (\ref{case903}) straightforwardly.

\section{Conclusion}
In this paper, we have studied the linearization of two coupled
SNODEs. In particular, we have introduced a new method of deriving
linearizing transformations from the first integrals for the given
equation.  The procedure is simple and straightforward.  From our
analysis we have demonstrated that one can have two wider classes of
linearizing transformations, namely Class-A and Class-B,  depending 
upon the nature of the independent
variables. In Class-A category, the independent variables are the same, and we identified three types of linearizing
transformations in which two of them are new to the literature.
On the other hand, in the Class-B category (the independent variables
are different), we found six new types of linearizing transformations. 
We have explicitly demonstrated the method of deducing the linearizing 
transformations and the general solution for all of these cases with 
specific examples. However, in this paper
we have restricted our attention on two aspects: (1) dependent variables
are functions of $(t,x,y)$ only, and (2) independent variables are not of
the local form $z_i=f_i(t,x,y,\dot{x},\dot{y})$, $i=1,2$. Linearization under 
these two types require separate treatment and will be studied subsequently. 
The method proposed here can naturally be extended to any number of coupled 
second order ODEs and indeed one can derive a very wide class of linearizing
transformations in these cases.

The work of MS forms part of a research project sponsored by National Board for Higher Mathematics, Government of India.  The work of ML forms part of a Department of Science and Technology, Government of India sponsored research project and is supported by a DST Ramanna Fellowship.

\end{document}